\documentclass[12pt]{article}
\begin{document}

\newcommand{\p}{\partial}
\newcommand{\be}{\begin{equation}}
\newcommand{\ee}{\end{equation}}
\newcommand{\del}{\bigtriangledown}
\newcommand{\G}{\Gamma}
\newcommand{\g}{\sqrt{-g}}
\newcommand{\half}{\frac{1}{2}}
\newcommand{\chris}{\left\{ \!\!\!\!\!\! \begin{array}{c} \lambda \\ 
\begin{array}{cc} \mu & \nu \end{array} \end{array} \!\!\!\!\!\!
\right\}}
\newcommand{\chrisa}{\left\{ \ \  \right\}}

{\hfill   November 1997 }

{\hfill    WATPHYS-TH97/17}

\vspace*{2cm}
\begin{center}
{\Large\bf Palatini Variational Principle for 
an Extended Einstein-Hilbert Action}
\end{center}

\begin{center}
{\large\bf Howard Burton\footnote{e-mail: hsburton@avatar.uwaterloo.ca}
and
Robert B.~Mann\footnote{e-mail: mann@avatar.uwaterloo.ca}} 
\end{center}
\begin{center}
{\it Department of Physics, University of Waterloo, Waterloo, Ontario 
N2L 3G1, Canada}  
\end{center}

\vspace*{1cm}
\begin{abstract}
\noindent
We consider a Palatini variation on a generalized Einstein-Hilbert action.
We find that the Hilbert constraint, that the connection equals
the Christoffel symbol, arises only as a special
case of this general action, while for particular values of the coefficients
of this generalized action, the connection is completely unconstrained.
We discuss the relationship between this situation and that usually
encountered in the Palatini formulation.
\end{abstract}
\newpage
\baselineskip=.8cm

\section{Introduction}

{}From the earliest days since the advent of General Relativity, attempts have
been made to generalize it.  The original motivations for doing so were concerned
with unifying gravitation and electromagnetism, which today have been superseded with
the desire to construct a theory of quantum gravity.  There are presently
many attempts to this end, including the superstring-theoretic \cite{str}
the connection dynamics proposal \cite{Ashtekar}, non-commutative geometries 
\cite{Madore}, Chern-Simons formulations \cite{CS},
gauge-theoretic formulations \cite{Hehl}, quantization 
of topologies \cite{Isham}, topological geons \cite{Sorkin},
gravity as an induced phenonemon \cite{Frolov}, and so on. 

Throughout this history the Palatini variational principle has played a subtle
but important role. As is well known, if one subjects the ordinary 
$N$-dimensional Einstein-Hilbert (EH) action,
\be\label{1}
S_{EH} = \int d^{N} x \left[ \g \left( R(\Gamma) + 16\pi\mathcal{L}_{m} 
\right) \right]
\ee
\medskip
to a Palatini variation, {\it i.e.} assumes that there is no {\it a-priori}
relationship between the (torsion-free) affine connection 
$\Gamma^\alpha_{\mu\nu}$ and the metric,  and thus subjects the action 
to a variation
$\delta_{\G} S = 0$ as well as $\delta_{g} S = 0$, one finds, in addition to 
the usual field equation resulting from the metric variation,
\be\label{2}
8\pi T_{\mu\nu} = G_{\mu\nu}(\G),
\ee
\medskip
from the  connection variation the constraint
\be\label{3}
\partial_\lambda g_{\mu\nu} -
\G^\eta_{\;\:\lambda\mu} g_{\eta\nu} -
\G^\eta_{\;\:\lambda\nu} g_{\mu\eta}  = 0
\ee
which is the familiar condition of metric compatibility, whose solution
\be\label{4}
\G^{\eta}_{\mu\nu} = \left\{ \!\!\!\!\!\! \begin{array}{c} \eta \\ 
\begin{array}{cc} \mu & \nu \end{array} \end{array} \!\!\!\!\!\!
\right\}
\ee
\medskip
is the Christoffel symbol.  In other words 
the geometrical constraint (\ref{3}) (henceforth called the "Hilbert constraint") 
is now a field equation that extremizes the action (\ref{1}). 
The fact that this seemingly independent line of inquiry corroborated the 
metrically compatible choice of the Christoffel symbol 
has been viewed by many as a kind of "proof" of the 
validity of the Hilbert (or 2nd-order) variation of the EH action, in which (\ref{3})
is given and (\ref{1}) is therefore a functional only of the metric degrees
of freedom.  Certainly it alters the Lagrangian formulation of general relativity 
insofar as it removes the need to include a boundary term because there are no
derivatives of field variations on the boundary \cite{Wald}.

However as noted by Schr\"{o}dinger long ago \cite{Sch}, and
emphasized by Hehl \cite{Hehl}, in a generalized theory of gravitation one
expects the geometrical relationship (\ref{3}) to be modified in some
manner that is typically not obvious. Hence the 2nd order variation is 
often not available, and one must resort to a Palatini-type of variational
principle. Indeed, the Palatini approach has been employed in most of the 
generalized theories of (quantum) gravity mentioned above, either in terms of
affine connection -- metric variables or (as is common in supergravity theories
\cite{supergrav}) spin connection -- vielbien variables.  Furthermore,
although the physical relevance of the metrically compatible 
Christoffel symbol in general relativity is clear, from a geometrical perspective
the singling out of the Christoffel connection is somewhat curious because 
the geometry is impervious to which particular connection is  
chosen (Christoffel or otherwise), as long as it is torsion-free.

Motivated by the above, we consider in this paper the relationship between the
Palatini variational principle and the condition of metric compatibility. Since
the key premise of the Palatini principle is that metric and connection
are independent of one another at the outset, we consider a generalization
of the EH action (\ref{1}) which includes all possible terms that are at
most quadratic in derivatives and/or connection variables. We then 
determine the circumstances under which a Palatini variational principle
yields the compatibility condition (\ref{3}), and what
the consequent gravitational dynamics would be in situations that are
more general.  We work in $N$ dimensions, and consider actions which
are functionals only of the metric and the affine connection (although
our approach could straightforwardly be extended to a vielbein formalism). 
For simplicity we consider only torsion-free connections.

\section{Generalized Action and Connection Constraints}

If one assumes that metric and connection variables are independent of one
another, then there is no longer any {\it a-priori} reason to consider
the EH action (\ref{1}) as the action on which to base a theory of gravitation.
One is guided only by principles of general covariance, minimal coupling,
simplicity, and logical economy.  

Hence we seek a Lagrangian which is a scalar under general coordinate transformations
and which has the minimal number of derivatives and/or powers of the field
variables in every term.  Since the connection does not transform like a tensor,
one must construct objects from it which have tensorial properties.  The simplest
of these are the Riemann curvature tensor 
\be\label{5}
R^\alpha_{\;\:\beta\mu\nu} = \partial_\nu\G^\alpha_{\;\:\beta\mu}
-\partial_\mu\G^\alpha_{\;\:\beta\nu} + \G^\alpha_{\;\:\sigma\mu}
\G^\sigma_{\;\:\beta\nu}
- \G^\alpha_{\;\:\sigma\nu} \G^\sigma_{\;\:\beta\mu}
\ee
and the covariant derivative of the metric
\be\label{6}
\del_\lambda g_{\mu\nu} = \partial_\lambda g_{\mu\nu} -
\G^\eta_{\;\:\lambda\mu} g_{\eta\nu} -
\G^\eta_{\;\:\lambda\nu} g_{\mu\eta} 
\ee
where in (\ref{5}) and (\ref{6}) the connection is assumed to be torsion-free.

The most general action in $N$ dimensions that one can construct out of these objects 
subject to these constraints is
\begin{eqnarray}
S & = & \int d^{N} x \g 
[R + H (\del_{\nu} g^{\alpha\beta})(\del^{\nu}g_{\alpha\beta}) + I V^{2}
+ J (\del_{\epsilon}g_{\mu\nu})(\del^{\mu}g^{\epsilon\nu}) \nonumber \\ 
&   & \hspace{1.8in} + K V \cdot Z + L Z\cdot Z], \label{7}
\end{eqnarray}
\medskip
where 
\be\label{8}
V_{\rho} := \frac{\del_{\rho}\g}{\g}
\qquad Z^{\lambda} := \del_{\eta} g^{\eta\lambda}
\ee
and where the coefficients $H$, $I$, $J$, $K$ and $L$ are constants.  
Other scalar quantities exist, but they either can be rewritten as
linear combinations of the terms in (\ref{7}) up to total derivatives 
or they are at least cubic in derivatives and/or connection variables.
Since we assume $\delta g_{\mu\nu}$ and $\delta\Gamma^\alpha_{\mu\nu}$
to vanish at the boundary, no additional boundary terms in (\ref{7})
are required.

Variation of (1) with respect to the connection $\G^{\lambda}_{\rho\sigma}$
leads to the following constraint
\begin{eqnarray}
\frac{1}{\g} \left( \del_{\lambda} \left[
\g g^{\rho\sigma} \right] - \half \del_{\epsilon} \left[
\g g^{\rho\epsilon} \right] \delta^{\sigma}_{\lambda} -
\half \del_{\epsilon} \left[
\g g^{\sigma\epsilon} \right] \delta^{\rho}_{\lambda} \right)  &  & \nonumber \\
+ H \left[ \left(
\del^{\rho}g^{\sigma\gamma} +
\del^{\sigma}g^{\rho\gamma}\right)g_{\gamma\lambda} -
\del^{\rho}g_{\lambda\gamma} g^{\sigma\gamma} -
\del^{\sigma}g_{\lambda\gamma} g^{\rho\gamma} \right]          &  &  \label{9}\\
+ I \left[ V^{\rho} \delta^{\sigma}_{\lambda} + 
V^{\sigma} \delta^{\rho}_{\lambda} \right]
+J \left[ g_{\nu\lambda} \left( \del^{\rho} g^{\sigma\nu} \right) +
\del_{\lambda}g^{\rho\sigma} - g^{\mu\rho} \left\{
g^{\nu\sigma} \left( \del_{\lambda} g_{\mu\nu} \right) +
\del^{\sigma} g_{\mu\lambda} \right\} \right]                   &  & \nonumber \\
+K \left[ \half \left( Z^{\sigma} \delta^{\rho}_{\lambda} + 
Z^{\rho} \delta^{\sigma}_{\lambda} \right) - \half \left(
V^{\sigma} \delta^{\rho}_{\lambda} + V^{\rho} \delta^{\sigma}_{\lambda} \right) -
V_{\lambda} g^{\rho\sigma} \right] -
L \left[ Z^{\rho} \delta^{\sigma}_{\lambda} + Z^{\sigma} \delta^{\rho}_{\lambda} +
2Z_{\lambda} g^{\sigma\rho} \right] & = &  0 \nonumber  
\end{eqnarray}
\medskip
whose solution determines the connection as a function of the metric in a
manner which generalizes (\ref{4}). 

We next seek to find the conditions under which (\ref{9}) may be solved
for $\G$ in terms of the metric.
Tracing (\ref{9}) on the $(\rho,\sigma)$ indices yields
\be\label{10}
\left[ (N-3)+2I-4J-(N+1)K \right] V_{\lambda} + \left[
4H+2J+K-2L(N+1)-1 \right] Z_{\lambda} = 0,
\ee
\medskip
whilst a $\rho-\lambda$ contraction of (\ref{9}) gives
\begin{eqnarray}\label{11}
 \left[(N-1)+8H-2(N+1)I+4J+(N+3)K \right] V_{\lambda}\qquad\qquad\qquad && \nonumber\\
\qquad + \left[(N-1)-4H-6J-(N+1)K+2(N+3)L \right] Z_{\lambda} &=& 0 .
\end{eqnarray}

Equations (\ref{10}) and (\ref{11}) are two equations in the two unknown vector
fields $V_\lambda$ and $Z_\lambda$. Provided the determinant of coefficients
is non-zero, the
only possible simultaneous solutions of (\ref{10}) and (\ref{11}) are
\be\label{12}
V_{\lambda} = Z_{\lambda} = 0 
\ee
\medskip
which implies 
\be\label{13}
-\del_{\lambda}g^{\rho\sigma} \left[ 1+2J \right] +
(2H+J) \left[
g_{\lambda\gamma} \left(\del^{\rho}g^{\sigma\gamma} 
+ \del^{\sigma}g^{\rho\gamma} \right) \right] = 0
\ee
upon insertion of (\ref{12}) into (\ref{9}).  It is straightforward to show that
\be\label{14}
\G^{\alpha}_{\mu\nu} = \frac{1}{2}g^{\alpha\lambda}\left(\partial_\mu g_{\alpha\nu} 
 + \partial_\nu g_{\alpha\mu} - \partial_\alpha g_{\mu\nu} \right)
\equiv  \left\{ \!\!\!\!\!\! \begin{array}{c} \eta \\ 
\begin{array}{cc} \mu & \nu \end{array} \end{array} \!\!\!\!\!\!
\right\}
\ee
is the only solution to (\ref{13}) provided that $3J+2H\neq -1$ or $H\neq \frac{1}{4}$. 
Consequently we see that metric compatibility arises within the Palatini formalism
under quite general conditions unless $3J+2H=-1$, in which case, for $J
\neq - \half$, it can be shown that
$\G^{\lambda}_{\;\:\mu\nu}$ is of the form:
\be
\G^{\lambda}_{\;\:\mu\nu} = \chris + g^{\lambda\gamma} \left[
\Upsilon_{\mu\gamma\nu} + \Upsilon_{\nu\gamma\mu}
-2\Upsilon_{\mu\nu\gamma}\right]
\ee
where $\Upsilon_{\mu\nu\gamma}$ is a tensor obeying 
$\Upsilon_{\mu\nu\gamma} = \Upsilon_{\nu\mu\gamma}$ and 
$g^{\mu\nu}\Upsilon_{\mu\nu\gamma} = g^{\mu\nu}\Upsilon_{\mu\gamma\nu}$ 
but is otherwise arbitary. Similarly if 
$H= \frac{1}{4}$ we find, again for $J \neq -\half$, that 
$\G^{\lambda}_{\;\:\mu\nu}$ is of the form:
\be
\G^{\lambda}_{\;\:\mu\nu} = \chris + g^{\lambda\gamma} \left[
{\Lambda}_{\mu\nu\gamma} + {\Lambda}_{\nu\gamma\mu} + 
{\Lambda}_{\gamma\mu\nu} \right]
\ee
where $\Lambda_{\mu\nu\gamma} = \Lambda_{\nu\mu\gamma}$
is an arbitrary tensor that is traceless on all indices.
We further note that the condition that trivializes
(\ref{13}), i.e. $J = -\half$, $H = \frac{1}{4}$, is a simultaneous 
solution of both of the above special cases and thus  
leaves $\del_{\lambda}g^{\rho\sigma}$ completely undetermined modulo the
conditions given in (\ref{12}).   In this case, the Palatini variation 
provides almost no information about the relationship between the
metric and the connection, as (\ref{12}) furnishes only 8 equations to
determine the 24 unknowns $\G$. Furthermore, eq. (\ref{12}) would not exist
if the determinant of coefficients in (\ref{10},\ref{11}) were set to
zero, thereby yielding a redundancy.

We expect that this redundancy is made manifest
by some symmetry on the connection coefficients.
To this end, consider the following general transformation of the
connection:
\be\label{15}
\G^{\lambda}_{\;\:\mu\nu} \Rightarrow \hat{\G}^{\lambda}_{\;\:\mu\nu}
= \G^{\lambda}_{\;\:\mu\nu} + Q^{\lambda}_{\;\:\mu\nu},
\ee
\medskip
where $Q^{\lambda}_{\;\:\mu\nu}$ is an arbitrary tensor field with the
sole restriction that,
like $\G^{\lambda}_{\;\:\mu\nu}$, it is symmetric in its last two indices.
This
type of transformation is sometimes called a deformation transformation
\cite{Hehl}.  Under the above transformation we find that the action (\ref{7}) 
is correspondingly transformed
\be\label{16}
S \Rightarrow \hat{S} = S + \delta S,
\ee
\medskip
where
\begin{eqnarray}
\delta S & = & -[1+2J](\del^{\lambda}g^{\mu\nu})Q_{\lambda\mu\nu}
	       -[2H+J](\del^{\lambda}g^{\mu\nu})(Q_{\mu\lambda\nu} + Q_{\nu\lambda\mu}) \nonumber \\
	 &   & -[1+2H+3J]Q^{\lambda\mu\nu}Q_{\nu\mu\lambda} - [2H+J]Q^{\lambda\mu\nu}Q_{\lambda\mu\nu} \nonumber \\
	 &   & +[I-K+L]Q_{\lambda}^{\;\:\lambda\rho}Q^{\epsilon}_{\;\:\epsilon\rho} 
+[1-K+2L]Q^{\lambda}_{\;\:\lambda\rho}Q^{\rho\epsilon}_{\;\;\;\epsilon}
\nonumber \\
	 &   &
+LQ_{\rho\epsilon}^{\;\;\;\epsilon}Q^{\rho\lambda}_{\;\;\;\lambda}
+ [1-2I+K]V_{\lambda}Q_{\epsilon}^{\;\:\epsilon\lambda} 
	       +[K-1]V_{\lambda}Q^{\lambda\eta}_{\;\;\;\eta}  \nonumber \\ 
	 &   & +2LZ^{\lambda}Q_{\lambda\eta}^{\;\;\;\eta} +
[1+2L-K]Z^{\lambda}Q^{\eta}_{\;\:\eta\lambda}
\label{17}
\end{eqnarray}
For $\G^{\lambda}_{\;\:\mu\nu}$ to be completely unconstrained, we must
have
$\delta S = 0$ regardless of the choice of $Q^{\lambda}_{\;\:\mu\nu}
(Q_{\lambda\mu\nu})$.  This 
can only happen 
\be\label{18}
H=\frac{1}{4};J=-\half;I=K=1;L=0,
\ee
\medskip
which we note ensures
that the determinant of coefficients in the system (\ref{10},\ref{11}) vanishes.

Conversely, consider subsitution of (\ref{15}) for
$\G^{\lambda}_{\;\:\mu\nu}$, 
into the general action (\ref{7}), and then varying the (transformed) action with 
respect to $Q^{\lambda}_{\;\:\mu\nu}$.  This yields a set of complicated
algebraic
equations for  $Q^{\lambda}_{\;\:\mu\nu}$. Insertion into (\ref{7}) of
their solution for
$Q^{\lambda}_{\;\:\mu\nu}$ in terms of $\G^{\lambda}_{\;\:\mu\nu}$ and
$g_{\mu\nu}$
leads directly to a modified action of the form given in (\ref{7}) whose
specific values for $H,I,J,K,L$ are given by (\ref{18}) above.\footnote{see 
Appendix for more explicit details} 

In other words, (\ref{18}) is clearly the unique set of values such that 
our action is invariant under the transformation (\ref{15})
with $Q^{\lambda}_{\;\:\mu\nu}$ completely unconstrained other than being 
symmetric in its lower two indices.  Accordingly, the values (\ref{18}) will
henceforth be called the "maximally symmetric" values.\footnote{``maximally" 
symmetric to distinguish them from other partial symmetries which may occur
when one assumes some particular tensorial structure in
$Q^{\lambda}_{\;\:\mu\nu}$.}

{}From this perspective one can say that the compatibility condition (\ref{3}), 
obtained by applying the
Palatini variational principle to the EH action, is an example of a constraint
induced by a broken symmetry.  That is, the EH action is a special case of our 
general action (\ref{7}) above, with the particular requirement that $H=I=J=K=L=0$.  
That these values of $H,I,J,K,L$ break the general symmetry is obvious from 
the above analaysis, and it is this breaking of this  "connection symmetry" which
singles out the Christoffel symbol.

\section{Extended Action Dynamics}

Momentarily putting aside our consideration of the ``connection-dynamics" 
of our extended action and calculating the ordinary ``metric-dynamics", we find
\begin{eqnarray}\label{19}
8\pi T_{\mu\nu} & = &  G_{(\mu\nu)}(\G) + (I-K)V_{\mu}V_{\nu}   
-\half K \left[ \del_{\mu}V_{\nu} + \del_{\nu}V_{\mu} \right] \nonumber \\
&   & -2 \left( \del_{\lambda} + V_{\lambda} \right) g^{\lambda\epsilon}
\left[ H \del_{\epsilon}g_{\mu\nu} + \half J \left(
\del_{\mu}g_{\nu\epsilon} + \del_{\nu}g_{\mu\epsilon} \right) \right] \nonumber \\
&   & -L \left[
V_{\mu}Z_{\nu} + V{\nu}Z_{\mu} + \del_{\mu}V_{\nu} + \del_{\nu}V_{\mu} + 
Z_{\mu}Z_{\nu} \right] \nonumber \\
&   & +H \left[
(\del_{\mu}g^{\alpha\beta})(\del_{\nu}g_{\alpha\beta}) + 2g^{\alpha\beta}(\del^{\lambda}g_{\alpha\mu})(\del_{\lambda}g_{\beta\nu}) 
\right] \nonumber \\
&   & + \half J \left[
(\del^{\eta}g_{\alpha\mu})(\del^{\alpha}g_{\nu\eta}) + (\del^{\eta}g_{\alpha\nu})(\del^{\alpha}g_{\mu\eta}) \right] \nonumber \\
&   & + g_{\mu\nu} \{ -\half H (\del_{\rho}g^{\alpha\beta})(\del^{\rho}g_{\alpha\beta}) 
+ \half I V^{2} 
- \half J (\del_{\epsilon}g_{\alpha\beta})(\del^{\alpha}g^{\epsilon\beta}) \nonumber \\
&   & \hspace{.5in} -\half L Z^{2} + \del_{\epsilon} \left(
IV^{\epsilon} + \half K Z^{\epsilon} \right) \} 
\end{eqnarray}
upon variation of (\ref{7}) with respect to the metric.  Provided the constants
$H,I,J,K,L$ are chosen so that (\ref{14}) is satisfied (i.e. the coefficients are  
chosen so that $3J+2H \neq -1$ and $H \neq \frac{1}{4}$), then all terms on the right hand
side of (\ref{19}) vanish except for the first one, which becomes the usual
expression for the Einstein tensor in terms of the metric.

Consider next the condition of maximal symmetry. 
Insertion of our maximally symmetric values, (\ref{18}), into the above dynamical
equation yields
\begin{eqnarray}\label{20}
8\pi T_{\mu\nu} & = & G_{(\mu\nu)}(\Gamma) + \frac{1}{4} \left[
\del_{\mu}P_{\nu\eta}^{\;\;\;\eta} + \del_{\nu}P_{\mu\eta}^{\;\;\;\eta} 
\right] + \left[ \del_{\lambda} - \half (P_{\lambda\eta}^{\;\;\;\eta}
\right]
(E^{\lambda}_{\;\:\mu\nu}) 
\nonumber \\
&   & + \frac{1}{4} \left[
2(P^{\lambda\beta}_{\;\;\;\mu})(P_{\lambda\beta\nu}) - (P_{\mu}^{\;\:\lambda\eta})(P_{\nu\lambda\eta})
-2(P^{\lambda}_{\;\:\eta\mu})(P^{\eta}_{\;\:\lambda\nu}) \right] \\
&   & + \frac{1}{8} g_{\mu\nu} \left[
2(P_{\lambda\eta}^{\;\;\;\eta})(P^{\lambda\rho}_{\;\;\;\rho})
-2(P_{\epsilon\lambda\eta})(P^{\lambda\epsilon\eta})
+(P_{\lambda\eta\epsilon})(P^{\lambda\eta\epsilon})
+4 \del_{\epsilon} \left(
P_{\lambda}^{\;\:\lambda\epsilon} - P^{\epsilon\eta}_{\;\;\;\eta} 
\right) \right] \nonumber 
\end{eqnarray}
where         
\be\label{21}
P_{\eta}^{\;\:\mu\nu} := \del_{\eta} g^{\mu\nu}
\ee
and
\be
E^{\lambda}_{\;\:\mu\nu} := \half \left[
P^{\lambda}_{\;\:\mu\nu} - P_{\mu\nu}^{\;\;\;\lambda} - P_{\nu\mu}^{\;\;\;\lambda} \right]
= \left[ \chris - \G^{\lambda}_{\;\:\mu\nu} \right],
\ee
\medskip
thus enabling us to put some terms directly in terms of the Christoffel symbol.

Hence the field equations in the case of maximal symmetry consist of (\ref{20})
alone -- there is no equation which determines the connection in terms of the metric.
In this sense the maximally symmetric action is a theory of gravity determined in
terms of metric dynamics alone, with the connection freely specifiable.

Since the connection may be freely specified, one choice is to make it compatible
with the metric, {\it i.e.} to demand that (\ref{14}) hold. In this case all
$P_{\eta}^{\;\:\mu\nu} = 0$, and (\ref{20}) reduces to
\be\label{23}
8\pi T_{\mu\nu}  =  G_{(\mu\nu)}(\chrisa)
\ee
which are the field equations for general relativity.  Alternatively, suppose we
choose $\Gamma^{\eta}_{\;\:\mu\nu} = 0$.  In this case (\ref{20}) becomes
\begin{eqnarray}\label{24}
8\pi T_{\mu\nu} & = & G_{(\mu\nu)}(\Gamma) + \frac{1}{4} \left[
\del_{\mu} \hat{P}_{\nu\eta}^{\;\;\;\eta} + \del_{\nu}\hat{P}_{\mu\eta}^{\;\;\;\eta} 
\right] + \left[ \del_{\lambda} - \half
\hat{P}_{\lambda\eta}^{\;\;\;\eta} \right] \chris 
\nonumber\\
&   & + \frac{1}{4} \left[
2(\hat{P}^{\lambda\beta}_{\;\;\;\mu})(\hat{P}_{\lambda\beta\nu}) - 
(\hat{P}_{\mu}^{\;\:\lambda\eta})(\hat{P}_{\nu\lambda\eta})
-2(\hat{P}^{\lambda}_{\;\:\eta\mu})(\hat{P}^{\eta}_{\;\:\lambda\nu})
\right] \\
&   & + \frac{1}{8} g_{\mu\nu} \left[
2(\hat{P}_{\lambda\eta}^{\;\;\;\eta})(\hat{P}^{\lambda\rho}_{\;\;\;\rho})
-2(\hat{P}_{\epsilon\lambda\eta})(\hat{P}^{\lambda\epsilon\eta})
+(\hat{P}_{\lambda\eta\epsilon})(\hat{P}^{\lambda\eta\epsilon})
+4 \del_{\epsilon} \left(
\hat{P}_{\lambda}^{\;\:\lambda\epsilon} - \hat{P}^{\epsilon\eta}_{\;\;\;\eta} 
\right) \right] \nonumber
\end{eqnarray}
where $\hat{P}_{\eta}^{\;\:\mu\nu} := \partial_{\eta} g^{\mu\nu}$.  Further simplification of
the right-hand side of (\ref{24}) yields
\be\label{25}
8\pi T_{\gamma\sigma} = G_{(\gamma\sigma)}(g)
\ee
where $G_{(\gamma\sigma)}(g)$ is the Einstein tensor expressed as a functional of
the metric, {\it i.e.} $G_{(\gamma\sigma)}(g) = 
G_{(\gamma\sigma)}(\chrisa)$. Hence (\ref{25}) also yields the equations of general
relativity.  

The above case of examining $\G = 0$ raises an interesting curiosity.  Clearly,
as the maximally symmetric case only restricts the connection to be torsion-free,
$\G = 0$ is an available option.  But the fact that we are able to choose such 
a connection \emph{globally} enables us to say something additional about the geometry 
of our manifold - namely that it is flat; or rather, that it can be made 
flat with no physical sacrifice.  

The preceding situation is also a generalization of a result obtained by 
Gegenberg {\it et. al.} for $(1+1)$ gravity \cite{Geg}. Consider the 
action (\ref{7}) for $N=2$ with
each of $H,I,J,K,L$ set to zero. In this case the determinant of coefficients
in eqs. (\ref{10}) and (\ref{11}) vanishes, and the general solution to (\ref{9})
is given by \cite{Geg}
\be\label{26}
\G^\alpha_{\mu\nu} = \bar{G}^\alpha_{\mu\nu} = \left\{ \!\!\!\!\!\! \begin{array}{c} \eta \\ 
\begin{array}{cc} \mu & \nu \end{array} \end{array} \!\!\!\!\!\!\right\}
+\left(\delta^\alpha_\mu B_\nu  + \delta^\alpha_\nu B_\mu - g_{\mu\nu} B^\alpha\right)
\ee
where $B_\mu$ is an arbitrary vector field. The Einstein tensor is given by
\begin{eqnarray}\label{19a}
G_{(\sigma\gamma)}(\bar{G}) & = & G_{(\sigma\gamma)}(\chrisa) \nonumber\\
& = & 0
\end{eqnarray}
and so renders the $(1+1)$ dimensional field equations trivial, as in the
usual Hilbert case.    We see from the preceding analysis of (\ref{20}) that
an analogous situation holds in higher dimensions for the maximally symmetric
action:  although the field equations do not determine the connection in terms of
the metric, one can choose the connection to be compatible with the metric by
appropriately choosing $Q^\alpha_{\;\:\mu\nu}$ in (\ref{15}) and recover
the
metric field equations of general relativity.

More generally, the choice of connection is completely irrelevant to the theory
in the maximally symmetric case.  One has only equation (\ref{20}), which determines
the evolution of the metric in terms of the basic matter fields.

\section{Conclusions}  

{}From the connection-dynamics perspective we have adopted in this paper, the
most general action which is 2nd order in connection and derivatives is given
by (\ref{7}).  In the usual formulation of the Palatini principle the $H,I,J,K,L$
coefficients are all set to zero.  We have shown that there exists a unique choice
of these coefficients, given by eq. (\ref{18}), 
such that the action is invariant under (\ref{15}). This case of maximal symmetry
yields a theory of gravity which is independent of the connection. 

{}From this
perspective the condition of metric compatibility (\ref{3}) in the usual Palatini
formulation arises as a field equation because this formulation breaks 
the maximal symmetry condition (\ref{18}), hence uniquely determining 
the connection. The equations of general relativity are recovered as 
a consequence of this broken symmetry.

In the maximally symmetric case we also recover the field equations of 
general relativity but for a different reason. In this case the connection
may be freely chosen by an appropriate choice of  
$Q^\alpha_{\;\:\mu\nu}$ in (\ref{15}), and so choosing it to be metrically
compatible obviously yields the metric field equations of general relativity. 
However these equations are recovered even if one does not choose the connection
to be compatible, as shown by the choice $\Gamma^\alpha_{\;\:\mu\nu} = 0$
in
the preceding section.  

Classically then, it would appear that maximally symmetric theories in the
Palatini formulation are classically equivalent to their broken counterparts,
at least insofar as metric dynamics is concerned. The role of maximally symmetric
theories in quantum gravity is, however, not clear, and would be interesting
to study further.

\section{Appendix}

The following is a proof of the claim made at the end of section 2:

If one begins with our usual generalized action, with H,I,J,K,L arbitrary,
that is:

\begin{eqnarray}
S & = & \int d^{N} x \g 
[R + H (\del_{\nu} g^{\alpha\beta})(\del^{\nu}g_{\alpha\beta}) + I V^{2}
+ J (\del_{\epsilon}g_{\mu\nu})(\del^{\mu}g^{\epsilon\nu}) \nonumber \\ 
&   & \hspace{1.8in} + K V \cdot Z + L Z\cdot Z], \label{7A}
\end{eqnarray}

and apply to it the variation:

\be\label{15A}
\G^{\lambda}_{\;\:\mu\nu} \Rightarrow \hat{\G}^{\lambda}_{\;\:\mu\nu} =
\G^{\lambda}_{\;\:\mu\nu} + Q^{\lambda}_{\;\:\mu\nu},
\ee

we find that $S$ consequently transforms to:

\be\label{16A}
S \Rightarrow \hat{S} = S + \delta S,
\ee

where

\begin{eqnarray}
\delta S & = & -[1+2J](\del^{\lambda}g^{\mu\nu})Q_{\lambda\mu\nu}
	       -[2H+J](\del^{\lambda}g^{\mu\nu})(Q_{\mu\lambda\nu} + Q_{\nu\lambda\mu}) \nonumber \\
	 &   & -[1+2H+3J]Q^{\lambda\mu\nu}Q_{\nu\mu\lambda} - [2H+J]Q^{\lambda\mu\nu}Q_{\lambda\mu\nu} \nonumber \\
	 &   &+[I-K+L]Q_{\lambda}^{\;\:\lambda\rho}Q^{\epsilon}_{\;\:\epsilon\rho} 
+[1-K+2L]Q^{\lambda}_{\;\:\lambda\rho}Q^{\rho\epsilon}_{\;\;\;\epsilon}
\nonumber \\
	 &   &
+LQ_{\rho\epsilon}^{\;\;\;\epsilon}Q^{\rho\lambda}_{\;\;\;\lambda}
+ [1-2I+K]V_{\lambda}Q_{\epsilon}^{\;\:\epsilon\lambda} 
	       +[K-1]V_{\lambda}Q^{\lambda\eta}_{\;\;\;\eta}  \nonumber \\ 
	 &   & +2LZ^{\lambda}Q_{\lambda\eta}^{\;\;\;\eta} +
[1+2L-K]Z^{\lambda}Q^{\eta}_{\;\:\eta\lambda}
\label{17A}
\end{eqnarray}

Now if we subject this new action, $\hat{S}$, to a variation with respect to
$Q^{\lambda}_{\;\:\alpha\beta}$, we clearly have:

\[ 
\delta_{\scriptscriptstyle Q^{\lambda}_{\;\:\alpha\beta}}\hat{S} = 
\delta_{\scriptscriptstyle Q^{\lambda}_{\;\:\alpha\beta}}(\delta S), \]
\medskip
since $\delta_{\scriptscriptstyle Q^{\lambda}_{\;\:\alpha\beta}}(S) = 0$.
\medskip
Now, from above we see that 
$\delta_{\scriptscriptstyle Q^{\lambda}_{\;\:\alpha\beta}}\hat{S} = 0 = 
\delta_{\scriptscriptstyle Q^{\lambda}_{\;\:\alpha\beta}}(\delta S)$
can be expressed as:
\begin{eqnarray}
\label{100}
0 & = & \int d^{N}x \g (\delta Q^{\lambda}_{\alpha\beta}) [ 
-(1+2J)\del_{\lambda}g^{\alpha\beta}
-(2H+J)g_{\lambda\mu}[\del^{\alpha}g^{\mu\beta}+\del^{\beta}g^{\mu\alpha}] \nonumber \\
&   & \hspace{1.4in} -[1+2H+3J](Q^{\alpha\beta}_{\;\;\;\lambda} + Q^{\beta\alpha}_{\;\;\;\lambda})
-2(2H+J)Q_{\lambda}^{\;\;\;\alpha\beta} \nonumber \\
&   & \hspace{1.4in} [I-K+L] \left[
(Q_{\epsilon}^{\;\:\epsilon\beta})\delta^{\alpha}_{\lambda} 
+ (Q_{\epsilon}^{\;\:\epsilon\alpha})\delta^{\beta}_{\lambda} \right]
\nonumber \\
&   & \hspace{1.4in} + [1-K+2L] \left( 
\half \left[
(Q^{\beta\epsilon}_{\;\;\;\epsilon})\delta^{\alpha}_{\lambda} +
(Q^{\alpha\epsilon}_{\;\;\;\epsilon})\delta^{\beta}_{\lambda} \right]
+Q^{\epsilon}_{\;\:\epsilon\lambda}g^{\alpha\beta} \right) \nonumber \\
&   & \hspace{1.4in} + g^{\alpha\beta} \left[
2L \left(
Q_{\lambda\epsilon}^{\;\;\;\epsilon} + Z_{\lambda} \right)  
+ (K-1)V_{\lambda} \right] \nonumber \\
&   & \hspace{1.4in} + \half \left[ 
(1-2I+K)V^{\beta} + (1+2L-K)Z^{\beta} \right] \delta^{\alpha}_{\lambda} \nonumber\\
&   & \hspace{1.4in} +
\half \left[
(1-2I+K)V^{\alpha} + (1+2L-K)Z^{\alpha} \right] \delta^{\beta}_{\lambda} ] 
\end{eqnarray}
\medskip			     
Clearly for arbitrary $\delta Q^{\lambda}_{\;\:\alpha\beta}$, we have 
the constraint that the coefficient in square brackets vanishes.
Taking the $g_{\alpha\beta}$ trace of this quantity yields
\be
\label{102}
AQ^{\epsilon}_{\;\:\epsilon\lambda} + B \left[ 
Q_{\lambda\epsilon}^{\;\;\;\;\epsilon} + Z_{\lambda} \right] + CV_{\lambda} = 0
\ee
while contracting over, say, $\lambda$ and $\alpha$ yields
\be
\label{103}
DQ^{\epsilon}_{\;\:\epsilon\lambda} + E \left[ 
Q_{\lambda\epsilon}^{\;\;\;\;\epsilon} + Z_{\lambda} \right] + FV_{\lambda} = 0
\ee
where
\be
A = \left[ (N-2)-4H+2I-6J-K(N+2)+2L(N+1) \right]
\ee
\be
B = \left[ 1-4H-2J-K+(1+N)L \right]
\ee
\be
C = \left[ (3-N)-2I+4J+(1+N)K \right]
\ee
\be
D = \left[ -6H+(N+1)I-5J-(N+2)K+(N+3)L \right]
\ee
\be
E = \left[ \half (N-1) -2H-3J- \half (N+1)K+(N+3)L \right]
\ee
\be
F = \left[ \half (N-1) +4H-(N+1)I+2J+ \half (N+3)K \right]
\ee
We note the following relationships:
\be
\label{104}
BD-AE = CE-BF
\ee
and
\be
\label{105}
F+D=E
\ee
Meanwhile, together (\ref{102}) and (\ref{103}) imply the following:
\be
\label{106}
[BD-AE]Q^{\epsilon}_{\;\:\epsilon\lambda} + [BF-CE]V_{\lambda} = 0
\ee
\medskip
Therefore, (\ref{104}),(\ref{105}) and (\ref{106}) in turn imply:
\be
\label{107}
Q^{\epsilon}_{\;\:\epsilon\lambda} = V_{\lambda}
\ee
and
\be
\label{108}
Q_{\lambda\epsilon}^{\;\;\;\epsilon} = - \left(
V_{\lambda} + Z_{\lambda} \right)
\ee
Inserting (\ref{107}) and (\ref{108}) into the coefficient of 
$\delta Q$ in (\ref{100}), yields, after a bit  
of symmetrization and manipulation:
\be
\label{109}
Q_{\lambda\alpha\beta} = \half \left[
P_{\lambda\alpha\beta} - P_{\alpha\beta\lambda} - P_{\beta\lambda\alpha} \right]
\ee
where
$P_{\mu}^{\;\:\nu\rho} = \del_{\mu}g^{\nu\rho}$
and $P_{\mu\nu\lambda} = -\del_{\mu}g_{\nu\lambda}.$
Inserting (\ref{107}),(\ref{108}) and (\ref{109}) into (\ref{17}) gives:
\be
\delta S = -S + \int d^{N}x \g \left[
R+ \frac{1}{4}(\del_{\alpha}g^{\mu\nu})(\del^{\alpha}g_{\mu\nu} - 
\half (\del_{\epsilon}g_{\mu\nu})(\del^{\mu}g^{\epsilon\nu}) + V^{2} + V \cdot Z
\right],
\ee
in other words, our maximally symmetric values.

\section*{Acknowledgements}
This work was supported in part by the Natural Sciences and Engineering Research
Council of Canada.  One of us (R.B.M.) would like to thank R. Sorkin and M. Grisaru
for interesting dicussions.

\end{document}